\documentclass[a4paper]{jpconf}

\usepackage{graphicx}

\newcommand{\be}{\begin{equation}}
\newcommand{\ee}{\end{equation}}
\newcommand{\bea}{\begin{eqnarray}}
\newcommand{\eea}{\end{eqnarray}}
\newcommand{\nn}{\nonumber}
\newcommand{\excleq}{\mathop{\stackrel{!}{=}}}
\newcommand{\ri}{\, {\rm i} \,}

\begin{document}

\title{Topological Summation in Lattice Gauge Theory}

\author{Wolfgang Bietenholz$^{1}$ and Ivan Hip$^{2}$} \

\address{$^{1}$ Instituto de Ciencias Nucleares, Universidad Nacional 
Aut\'{o}noma de M\'{e}xico
\\ ~\, A.P. 70-543, C.P. 04510 Distrito Federal, Mexico\\ \ \\
$^{2}$ Faculty of Geotechnical Engineering, University of Zagreb \\
~\, Hallerova aleja 7, 42000 Vara\v{z}din, Croatia }

~~\ead{wolbi@nucleares.unam.mx}

\begin{abstract}
In gauge theories the field configurations often occur in
distinct topological sectors. In a lattice regularised
system with chiral fermions, these sectors can be defined
by referring to the Atiyah-Singer Index Theorem. However, if such 
a model is simulated with local updates of the lattice gauge 
configuration, the Monte Carlo history tends to get stuck in one sector 
for many steps, in particular on fine lattices. Then expectation values 
can be measured only within specific sectors. Here we present a
pilot study in the 2-flavour Schwinger model which explores
methods of approximating the complete result for an observable ---
corresponding to a suitable sum over all sectors --- based
on numerical measurements in a few specific topological sectors.
We also probe various procedures for an indirect evaluation of
the topological susceptibility, starting from such topologically
restricted measurements.
\end{abstract}

\section{Topological sectors in gauge theories}
\

Our general framework in this article is the functional integral 
formulation of quantum physics in Euclidean space. In this setting,
the set of configurations may occur in disjoint subsets, so that
all continuously deformed configurations belong to the same subset.
Such subsets are known as {\em topological sectors.} 
Continuous deformations capture all configurations in
one topological sector, but none of any different sector
(general aspects are discussed {\it e.g.}\ in Refs.\ \cite{topo}).

The simplest example where this situation occurs is a quantum
mechanical scalar particle moving on the circle $S^{1}$, with 
periodic boundary conditions in Euclidean time. The expectation
value of some observable ${\cal O}$ in this system is given by
\be
\langle {\cal O} \rangle = \frac{1}{Z} \int {\cal D} \varphi \
{\cal O} [\phi ] \exp ( - S [\varphi ]) \ , \quad
{\rm where} \quad
Z = \int {\cal D} \varphi \ \exp ( - S [\varphi ])
\ee
is the partition function, and ${\cal D} \varphi$ is the sum over
all closed paths $\varphi (t) \in S^{1}$ in some period
$T$, {\it i.e.}\ $t \in [0,T]$ and $\varphi (0) = \varphi (T)$.
The set of all these paths is naturally divided into disjoint
subsets, which are characterised by the winding number
\be
Q = \frac{1}{2 \pi} \int_{0}^{T} dt \ \dot \varphi \in Z \!\!\! Z \ ,
\ee
which represents in this case the {\em topological charge}.
Continuous path deformations cannot change $Q$, hence these
subsets are indeed topological sectors. \\

Topological sectors also occur in a variety of gauge theories 
\cite{topo}. Let us consider gauge configurations in a Euclidean space 
with periodic boundary conditions (a torus). If they are split into
topological sectors, the characteristic topological charge $Q$
is also denoted as the Pontryagin index. Two examples are
\bea
{\rm 2d} \ U(1) &:& Q = \frac{1}{\pi} \int d^{2}x \,
\epsilon_{\mu \nu} F_{\mu \nu}(x) \ , \nn \\ 
{\rm 4d} \ SU(3) &:& Q = \frac{1}{32 \pi^{2}} \, {\rm Tr} \, 
\int d^{4}x \, F_{\mu \nu}(x) \, \tilde F_{\mu \nu}(x) 
\ ,
\label{topgauge}
\eea
where $F_{\mu \nu}$ is the field strength tensor, and
$\tilde F_{\mu \nu} := \epsilon_{\mu \nu \rho \sigma} F_{\rho \sigma}$.
Gauge configurations can be continuously deformed only within
a fixed topological sector, hence the functional integral splits into
separate integrals for each $Q \in Z \!\!\! Z$. \\

Let us now address such a gauge theory in the presence of {\em chiral
fermions,} {\it i.e.}\ massless fermions with a Dirac operator $D$
that anti-commutes with $\gamma_{5}$, $D\gamma_{5} + \gamma_{5} D = 0$.
In this case the zero modes of the Dirac operator have a definite
chirality $\pm 1$. 

For such a Dirac operator, in a given gauge background,
we denote the number of zero modes with chirality $+1 \ (-1)$
as $n_{+} \ (n_{-})$. Their difference is the {\em fermion index}
\be
\nu := n_{-} - n_{+} \in  Z \!\!\! Z \ .
\ee
The {\em Atiyah-Singer Index Theorem} \cite{AS} states that for any 
gauge configuration, this index coincides with the topological charge
\be
Q \excleq \nu \ .
\ee

\section{Lattice regularisation}
\

The lattice regularisation reduces the (Euclidean) space to
discrete sites $x$, which are separated by some finite
lattice spacing $a$. The latter implies an UV regularisation
of the corresponding quantum field theory.
Matter field variables are now defined on each lattice site,
{\it e.g.}\ $\bar \Psi_{x}, \ \Psi_{x}$ for fermion fields,
while gauge fields can be formulated as {\em compact link
variables} $U_{x,\mu} \in \{ \, {\rm gauge ~ group} \, \} \, $.
It is a great virtue that this formulation is {\em gauge invariant}
even on the regularised level, so in this approach no gauge
fixing is needed.

{\it A priori} there are {\em no} topological sectors anymore
in the lattice regularised system; all configurations can now
be continuously deformed into each other. Still, the desired
connection to the continuum theory motivates the attempt to
introduce somehow (the analogue of) topological sectors also
on the lattice. A number of suggestions appeared in the literature,
often with a somewhat questionable conceptual basis. A clean
formulation emerged only at the very end of the last century,
based on chiral lattice fermions. Their lattice Dirac operator $D$
cannot simply anti-commute with $\gamma_{5}$ due to the notorious
doubling problem of lattice fermions \cite{NN}, but it may obey the
{\em Ginsparg-Wilson Relation} (GWR), which reads (in its simplest form)
\be  \label{GWR}
D \gamma_{5} + \gamma_{5} D = a D \gamma_{5} D \ .
\ee
This still guarantees a lattice deformed --- but exact --- version of 
the chiral symmetry \cite{ML98}. The latter also implies that the
corresponding lattice Dirac operator has exact zero modes with
a definite chirality, as in the continuum. Hence we can adopt
the Index Theorem \cite{HLN} and define the topological charge of 
a lattice gauge configuration as $Q := \nu$.

We remark that random lattice gauge configurations always
occur with $n_{+}=0$ or $n_{-}=0$; configurations with a
cancellation in the lattice fermion index also exist (the free
fermion is an example), but their probability measure seems
to vanish.\footnote{The same holds for the specific lattice
field configurations which are exactly {\em on} a topological
boundary, so we can ignore them.}

\section{Monte Carlo simulation}
\

Observables in quantum gauge theory can be evaluated
beyond perturbation theory by means of Monte Carlo simulations.
The idea is to use a sizeable set of gauge configurations
$[U]$ (consisting of link variables all over the lattice volume), 
which are generated randomly with the probability distribution
\be  \label{fermidet}
p [U] = {\rm det} \, D[U] \ \exp (-S_{\rm gauge}[U]) \ . 
\ee
Here we assume a fermion action, which is bilinear
in the Grassmann valued spinor fields $\bar \Psi$, $\Psi$.
Their functional integration ${\cal D} \bar \Psi  {\cal D} \Psi$
is carried out already, giving rise to the
fermion determinant ${\rm det} \, D[U]$. 

The summation over this set of configurations yields a numerical
measurement of expectation values $\langle \dots \rangle$, in
particular of $n$-point functions. These results obviously come
with some statistical error (since the available set of 
configurations is finite), and a systematic error ({\it e.g.}\
due to the finite lattice spacing $a$, which usually
requires a continuum extrapolation). Both can be estimated and
reduced if necessary by extended simulations. On the other hand,
we stress again that the result is fully {\em non-perturbative;} we 
deal with the complete action in the exponent, {\it i.e.}\ we capture 
directly the given model at finite interaction strength. \\

Practical algorithms for the generation of gauge configurations 
(with the given probability distribution) perform local updates,
{\it i.e.}\ in one step a configuration is modified just locally. 
Iterating such steps many times leads to a 
(quasi-)independent new configuration, to be used for the
next measurement. Changing a gauge configuration
drastically in a single step is also conceivable in principle,
but in practice such algorithms tend to be inefficient.\footnote{Cluster 
algorithms are a counter example for certain spin models, but no efficient 
application to lattice gauge theories is known so far.}

A problem with a sequence of local updates is, however, that
it hardly ever changes the topological sector --- although
one should do so frequently in order to sample correctly the entire 
space of configurations. This problem is particularly striking 
in the attempts to simulate QCD with chiral quarks; the JLQCD
Collaboration performed very extensive 2-flavour QCD
simulations of this kind \cite{JLQCD} --- which led to 
interesting results --- but the Monte Carlo histories were always 
confined to the trivial topological sector of charge $Q=0$.

Most QCD simulations with dynamical quarks involve a non-chiral
lattice quark formulation, since Ginsparg-Wilson fermions
are tedious to simulate. In particular Wilson fermions
(and variants thereof) have the disadvantage of additive mass
renormalisation, but the problem with sampling the topological
sectors is less severe so far. However, that property depends 
on the lattice spacing; typical values that have been used
in the past are $a \approx 0.05 \dots 0.1 ~ {\rm fm}$. 
Once one tries to proceed to even finer lattices, the problem
of confinement of the Monte Carlo history to a single topological 
sector is expected to show up also in this formulation 
\cite{Lusch10}.\footnote{To be more explicit: any algorithm has
to obey ``detailed balance'', {\it i.e.}\ the transition probabilities
of some configuration $C_{1}$ to $C_{2}$ and vice versa have to match the
probability ratio for these configurations to occur (Boltzmann weights),
$p (C_{1} \to C_{2}) /p (C_{2} \to C_{1}) = 
\exp(S [C_{1}]-S [C_{2}])$.
The boundaries between
topological sectors are surrounded by zones of high action, {\it i.e.}\ 
low probability. As the lattice spacing $a$ is reduced, their weight 
$p(C)$ decreases with a high power of $a$ \cite{Lusch10}.
Hence a sequence of small update steps will rarely
tunnel through such a boundary.}

So we have to address the question how to handle
Monte Carlo simulations if the history tends to be trapped
for a very long (computing) time, {\it i.e.}\ for many, many
update steps, in {\em one} topological sector.
What are then the prospects for measuring some
$n$-point function, or the topological susceptibility
\be
\chi_{t} := \frac{1}{V} \left( \langle Q^{2} \rangle - \langle Q 
\rangle^{2} \right) \qquad (V = {\rm volume}) \ ,
\ee
which actually require the summation over a variety of topological
sectors, with suitable statistical weights?

This is a delicate and highly relevant issue.
Here we address it in a toy model study of the 2-flavour Schwinger
model, which we simulated \cite{BHSV} with {\em dynamical overlap 
hypercube fermions;} this is one version of chiral lattice
fermions \cite{ovHF}, with a Dirac operator that solves 
the GWR (\ref{GWR}).\footnote{In this formulation we insert an 
improved kernel into the overlap formula, instead of the Wilson
kernel of the standard overlap operator \cite{alt}. The virtues
include an improved locality and scaling behaviour, and approximate
rotation symmetry \cite{ovHF}.}
We designed and applied a variant of the Hybrid Monte Carlo
algorithm, which is particularly suitable for this type of
lattice fermions \cite{BHSV}.

\section{The Schwinger model}
\

The Schwinger model \cite{schwing} represents Quantum Electrodynamics 
on a plane (QED$_{2}$). It is a popular toy model; in particular it 
shares with QCD the property of fermion confinement \cite{CJS} (although
the gauge group is Abelian) and the presence of topological sectors,
see eq.\ (\ref{topgauge}). On the other hand there are qualitative
differences, such as the absence of a running gauge coupling in the
Schwinger model. In the continuum its Lagrangian can be written as
\be
{\cal L}(\bar \Psi , \Psi , A_{\mu}) = \bar \Psi (x) \Big[
\gamma_{\mu} ( \ri \partial_{\mu} + g A_{\mu}(x) ) + m \Big]
\Psi (x) + \frac{1}{2} F_{\mu \nu} (x) F_{\mu \nu} (x) \ .
\ee
We are interested in the case of $N_{f}=2$ degenerate fermion flavours
of mass $m \ll g$, where Ref.\ \cite{Smilga} made the following
predictions:
\bea
{\rm chiral~condensate} && \Sigma := - \langle \bar \Psi \Psi
\rangle = 0.388 \dots (m g^{2})^{1/3} \ , \label{Sigmatheo} \\
{\rm pion~mass} && M_{\pi} = 2.008 \dots (m^{2}g)^{1/3} \ .
\label{Mpitheo}
\eea
As in 2-flavour QCD, a ``meson'' singlet and a triplet
emerge, the former (latter) being massive (massless) in the
chiral limit $m \to 0$, cf.\ eq.\ (\ref{Mpitheo}). Referring
to this analogy, and in agreement with the literature, we
denote the triplet as ``pions''. Its emergence in 2 dimensions
might appear somewhat surprising; the theoretical background of 
these ``quasi-Nambu-Goldstone bosons'' was first discussed in 
Ref.\ \cite{Cole76}.

\section{Numerical measurement at fixed topology}
\

We simulated the 2-flavour Schwinger model at $\beta := 1/g^{2}=5$
\cite{BHSV}. This implies smooth gauge configurations (mean 
plaquette value $\simeq 0.9$). Also the ``meson'' dispersion
relations confirm that lattice artifacts are tiny \cite{BHSV}, hence 
we can confront our results directly with the continuum predictions
(\ref{Sigmatheo}), (\ref{Mpitheo}), without really needing a continuum 
extrapolation.
On the other hand, finite size effects are significant, and they
are in fact necessary for our discussion of topology dependent 
observables.
\begin{table}
\centering
\begin{tabular}{|c|c||r|r|r||r||r|}
\hline
$L$ & $m$ & \multicolumn{4}{|c||}{number of configurations} & 
topological \\
 &  & $\nu =0$   & $|\nu |= 1$ & $|\nu |= 2$ & 
total & transitions \\
\hline
\hline
16 & 0.01 & 2428  &  307 &     & 2735 &  7 \\  
\hline
16 & 0.03 & 1070  &  508 &     & 1578 &  2 \\
\hline
16 & 0.06 &  741  &  660 &     & 1401 &  7 \\ 
\hline
16 & 0.09 &  919  &  587 &   1 & 1507 &  7 \\
\hline
16 & 0.12 &  664  &  501 & 248 & 1413 &  8 \\
\hline
16 & 0.18 &  791  &  563 &  50 & 1404 & 15 \\
\hline
16 & 0.24 &  576  &  978 &  56 & 1637 & 17 \\
\hline
\hline
\hline
$L$ & $m$ & \multicolumn{4}{|c||}{number of configurations} & \\
    &     & $\nu =0$   & $|\nu |= 1$ &
$|\nu |= 2$ & $|\nu |= 3$ & total \\
\hline
\hline
20 & 0.01 & 435 & 304 &     &     & 739 \\
\hline
24 & 0.01 &     &     & 278 & 273 & 551  \\
\hline
28 & 0.01 &     & 240 &     & 180 & 420  \\
\hline
32 & 0.01 & 138 &  98 &  82 &     & 318 \\
\hline
\hline
32 & 0.06 &  91 & 293 &     &     & 384 \\
\hline
\end{tabular}
\caption{Statistics for lattice size $L=16$ (above) and
$L>16$ (below) and various fermion
masses $m$, in distinct topological sectors 
($\nu$ is the fermion index). 
Multiple starts of Monte Carlo histories were
necessary to get access to various topological sectors,
since topological transitions were very rare.}
\label{statab}
\end{table}
We simulated on $L \times L$ lattices of sizes
$L = 16, \, 20, \, 24, \, 28, \, 32$ with fermion masses
in the range $m = 0.01 \dots 0.24$ (in lattice units). 
Our statistics is displayed in Table \ref{statab}.\\

\begin{figure}[h!]
\begin{center}
\includegraphics[width=13pc,angle=0]{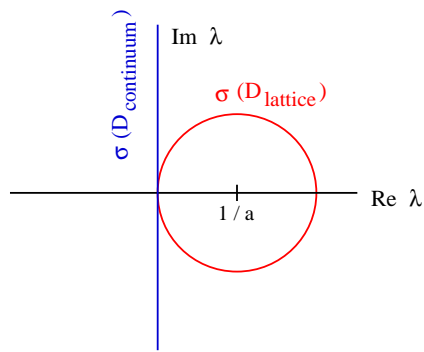}
\end{center}
\caption{\label{Dspec} The spectrum of a lattice Dirac operator that
fulfils the GWR (\ref{GWR}) is located on a
circle in the complex plane. In the continuum
limit $a \to 0$ it turns into the imaginary axis.}
\end{figure}

Let us first address the {\em Dirac spectrum.} All the eigenvalues
of a lattice Dirac operator (before adding the mass), which 
obeys the GWR (\ref{GWR}), are located on the circle in the
complex plane with centre and radius $1/a$, as illustrated in
Fig.\ \ref{Dspec}.
This confirms that the zero modes are exact, and we have
mentioned before that their fermion index is identified with the
topological charge, $Q = \nu = n_{-} - n_{+}$.\footnote{Actually
throughout this study only the absolute value $|\nu |$ matters.}

\begin{figure}[h!]
\begin{center}
\includegraphics[width=13pc,angle=270]{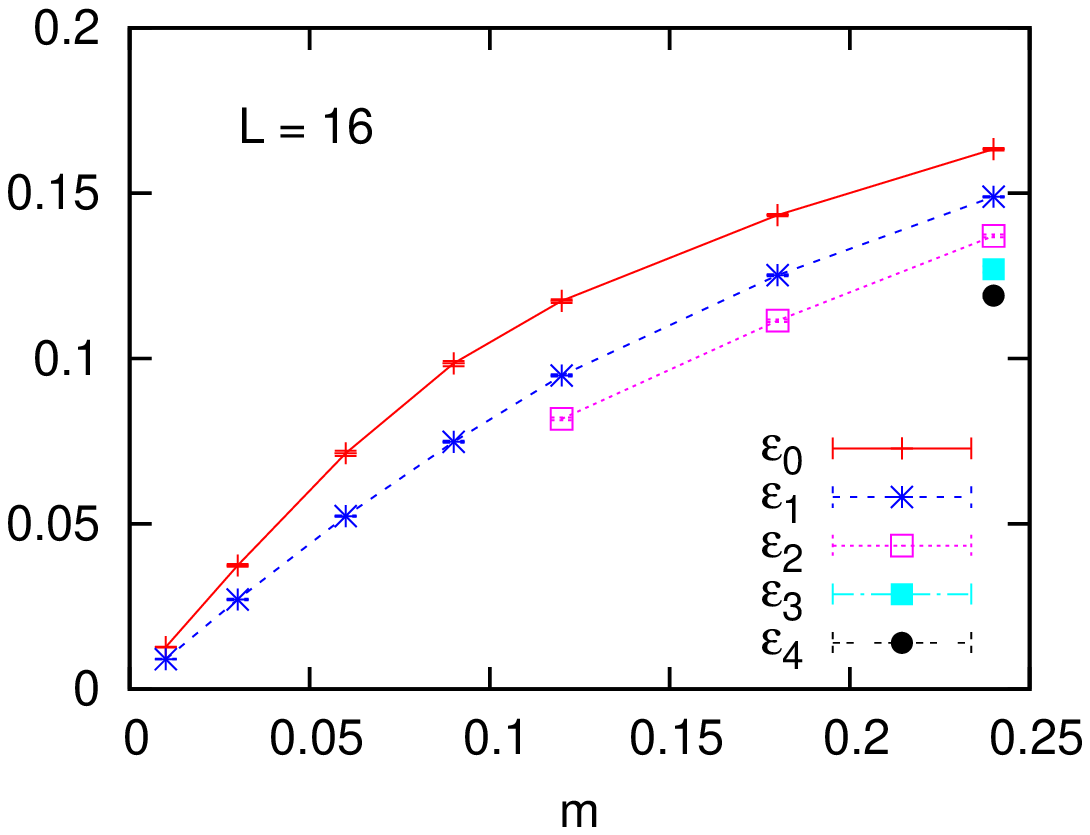}
\includegraphics[width=13pc,angle=270]{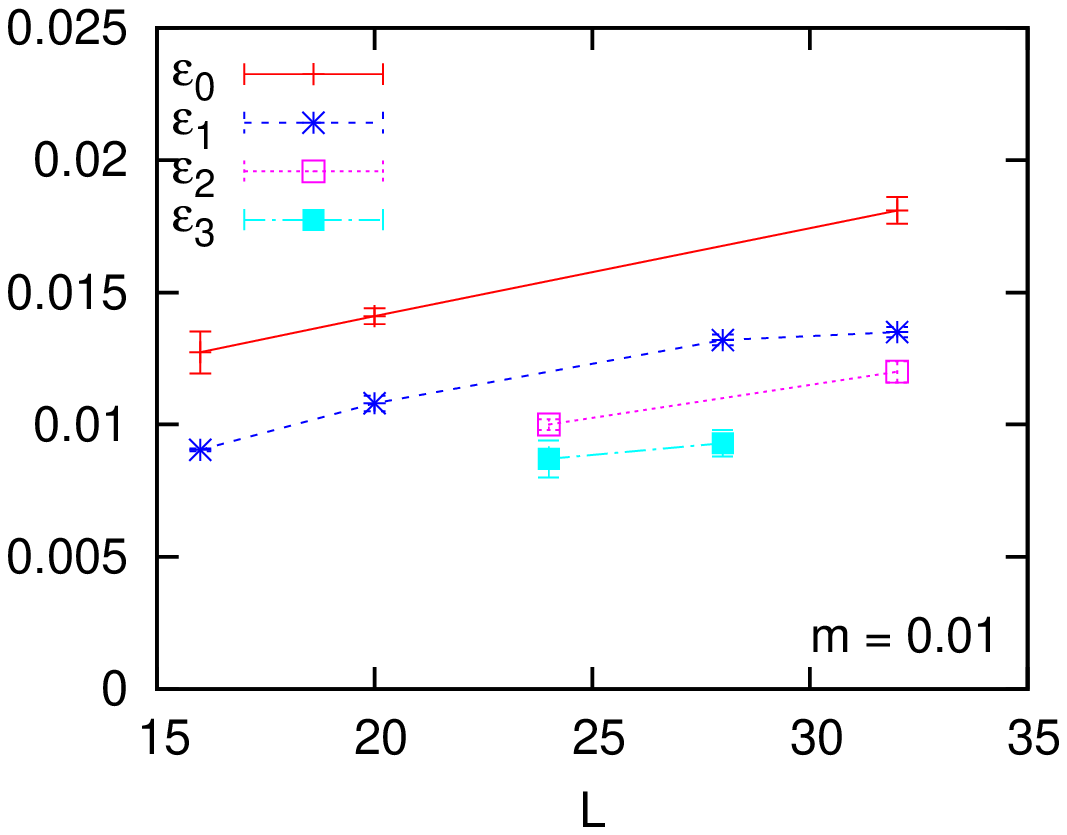}
\end{center}
\caption{\label{epsfig} Numerical results for the chiral condensate at
$|$topological charge$| = | \nu |$, after subtracting the zero mode
contribution, cf.\ eq.\ (\ref{Sigmanu}), 
at $L=16$ (on the left) and at $m=0.01$ (on the right).}
\end{figure}

In this study we could evaluate the complete Dirac spectrum for
our lattice configurations (which is not feasible in 4 dimensions,
except for tiny lattices). To make this spectrum compatible with the
continuum formulation, we map it stereographically onto the imaginary
axis \cite{BJS}. Based on the eigenvalues $\lambda_{i}$ that we obtain 
after this mapping, we obtain the {\em chiral condensate}
\be
\Sigma := - \langle \bar \Psi \Psi \rangle =  
\frac{1}{V} \left\langle \sum_{i} \frac{1}{|\lambda_{i}| + m} 
\right\rangle 
\ .
\ee
The sum can be computed for each configuration, but expectation values
can only be measured within fixed topological sectors. Table \ref{statab}
shows that topological transitions are indeed so rare that the entire
space of configurations is not well sampled, but specific sectors
are explored well. Hence we measure results for the expectation values 
of the chiral condensate at specific values of $|\nu |$,
\be
\Sigma_{|\nu |} = - \langle \bar \Psi \Psi \rangle_{|\nu|} 
= \frac{1}{V} \left\langle \sum_{i} 
\frac{1}{|\lambda_{i}| + m} \right\rangle_{|\nu|}
:= \frac{|\nu |}{mV} + \varepsilon_{|\nu|} \ .
\label{Sigmanu}
\ee
In the last expression we split off the zero mode contribution to 
$\Sigma_{|\nu |}$, which dominates at small mass $m$ (and $\nu \neq 0$), 
and we denote the rest as $\varepsilon_{|\nu|}$. Numerical results are 
shown in Fig.\ \ref{epsfig}.

It is a generic property of stochastic Hermitian matrices
(such as $\gamma_{5} D$) that zero eigenvalues repel the
low-lying non-zero modes. This suggests the inequality 
\be
\varepsilon_{0} > \varepsilon_{1} > \varepsilon_{2} \dots
\label{ineq1}
\ee
at fixed $m$ and $V$, which is confirmed consistently
by the plot in Fig.\ \ref{epsfig}
on the left. Moreover the plot on the right shows that
\be
\varepsilon_{i} (V_{1}) > \varepsilon_{i} (V_{2}) \quad {\rm for}
\quad V_{1} > V_{2} \ ,
\label{ineq2}
\ee
which is less obvious: in a larger volume more eigenvalues cluster
near zero, which supersedes the pre-factor $1/V$.

The rest of this article is devoted to tests of three different 
methods for approximately extracting ``physical'' quantities ({\it i.e.}\ 
quantities which are properly summed over all topological sectors),
based on measurements in a few specific sectors.

\section{Gaussian evaluation of the topological susceptibility}
\

We first assume a Gaussian distribution of
the topological charges --- this is certainly reasonable,
for instance precision tests in $SU(3)$ pure gauge theory
revealed at most tiny deviations from this behaviour \cite{topGaus}. 
It implies that the chiral condensate is composed as
\bea  \label{Sigmasum}
\Sigma &=& \sum_{\nu = - \infty}^{\infty} p(|\nu |) 
\ \Sigma_{|\nu |} \quad , \qquad 
p(|\nu |) = \frac{\exp \{ - \nu^{2} / (2 V \chi_{t})\} }
{\sum_{\nu} \exp \{ - \nu^{2} / (2 V \chi_{t})\} } \ .
\eea
Parity symmetry assures that $\langle \nu \rangle =0$, hence the 
topological susceptibility simplifies to
\be 
\chi_{t} = \frac{\langle \nu^{2}\rangle }{V} \ .
\ee
In most volumes we have data for $\Sigma_{0} \dots \Sigma_{Q}$,
{\it i.e.}\ up to some maximal topological charge $Q$. 
Thanks to inequality ({\ref{ineq1}) all the higher charge
contributions --- for $|\nu | > Q$ --- are bounded as
\be
\frac{|\nu |}{mV} < \Sigma_{|\nu |} < 
\frac{|\nu |}{mV} + \varepsilon_{Q} \ .
\ee
Hence for a given value of the susceptibility $\chi_{t}$
the sum in eq.\ (\ref{Sigmasum}) can be performed, up to a
uncertainty which affects $\Sigma$ only mildly, since
$\Sigma_{|\nu |}$ for high charges contribute only little.

In two volumes, $L=24$ and $28$, some $\Sigma_{|\nu |}$
data are missing for $|\nu | < Q$ (see Table \ref{statab}); 
in these cases we can again fix a minimal and a 
maximal value for $\varepsilon_{|\nu |}$,
this time based on inequality (\ref{ineq2}) and the results
in the next smaller and next larger volume.

So we can probe any ansatz for $\chi_{t}$ and compute the
corresponding value of $\Sigma$ up to a modest uncertainty.
We require the result to agree (within errors) with the prediction
(\ref{Sigmatheo}). In this way we determine $\chi_{t}$.
Fig.\ \ref{gausum} shows the results for $L=16$ and 
$m=0.01, \ 0.03, \ 0.06$ --- for higher masses the assumption
$m \ll g \simeq 0.45$, which is needed for the prediction
(\ref{Sigmatheo}), seems to fail. Since the theory refers to
infinite volume, we expect the result to improve for increasing $m$ 
({\it i.e.}\ for shorter correlation length) within the allowed range.
\begin{figure}[h!]
\begin{center}
\includegraphics[width=13pc,angle=270]{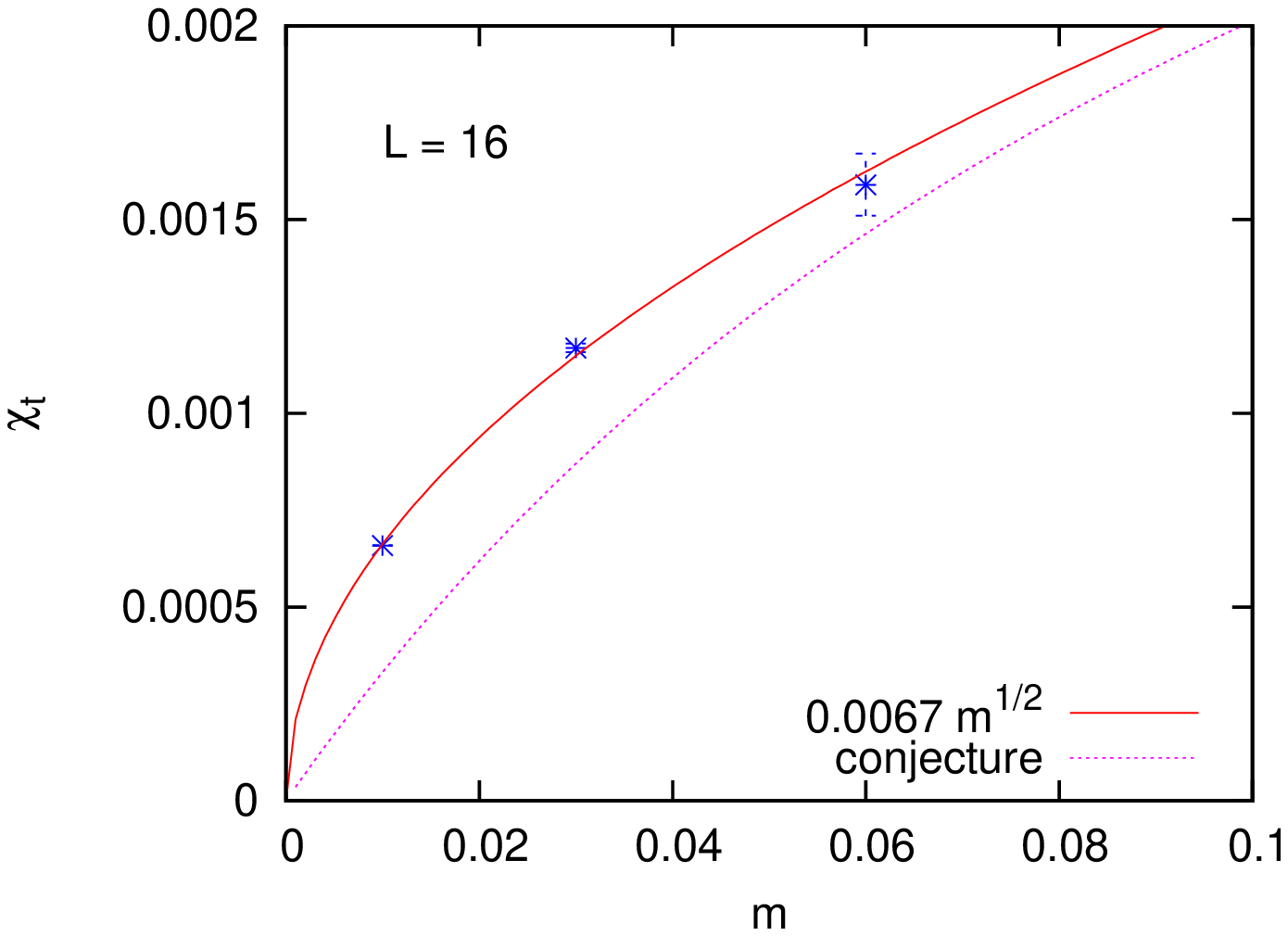}
\includegraphics[width=13pc,angle=270]{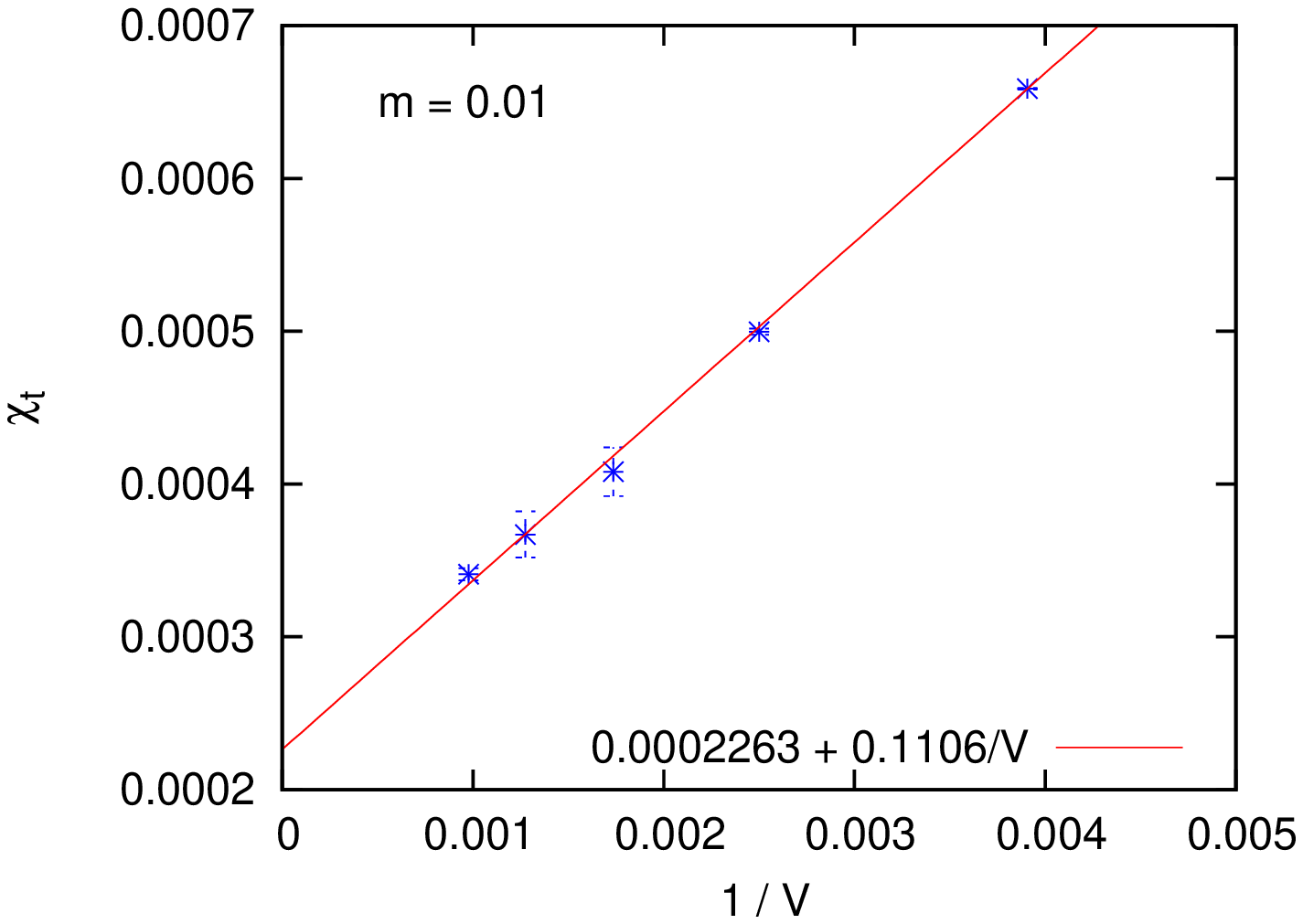}
\end{center}
\caption{\label{gausum} Our results for the topological 
susceptibility $\chi_{t}$ based on Gaussian summation.
On the left: in a fixed volume $V=16^{2}$ our data are compatible with 
an interpolation $\chi_{t}(m) \propto \sqrt{m}$, and for increasing
mass ({\it i.e.}\ reduced finite size effects) we approach
the conjecture (\ref{Durrconj}). On the right: 
at fixed $m=0.01$ we obtain results for $\chi_{t}$, which seem
to follow a behaviour linear in $1/V$.}
\end{figure}

The result is compared to a QCD-inspired conjectured of Ref.\ 
\cite{Durr} (for $N_{f}$ flavours, in a large volume),
\be  \label{Durrconj}
\frac{1}{\chi_{t}} = \frac{N_{f}}{m} \, \Sigma
(N_{f}=1, m=0) + \frac{1}{\chi_{t} (N_{f}=0)} \ .
\ee
The first ingredient has been computed analytically,
$\Sigma (N_{f}=1, m=0) \simeq 0.16 \, g$ \cite{Cole76},
and the quenched susceptibility $\chi_{t} (N_{f}=0) \simeq 0.000332$ 
has been measured numerically \cite{DuHo}. Fig.\ \ref{gausum}
confirms that the corresponding curve approaches the fit through our 
values for increasing $m$.

Alternatively we fix the mass $m=0.01$ and compare the results
for $L= 16 \dots 32$ (plot in Fig.\ \ref{gausum} on the right).
In our largest volume, $V=32^{2}$, we obtained $\chi_{t} =
0.000341(4)$, which is close to the value given by
conjecture (\ref{Durrconj}), $\chi_{t}= 0.000332$. An infinite volume
extrapolation of our data, however, leads to a smaller susceptibility 
of $\chi_{t} = 0.000226(5)$.

\section{Correlation of the topological charge density}
\

A drawback of the method in Section 6 is
that a known reference quantity is needed (here it was $\Sigma$),
and results in various topological sectors are required.
This is not the case for an approach suggested in Ref.\ 
\cite{AFHO}, which derived a ``model independent formula''
for the {\em correlation of the topological charge density} 
$\rho_{t}$ in one sector,
\be  \label{densecorr}
^{~\lim}_{|x|\to \infty} \ \langle \rho_{t}(x) \rho_{t}(0)
\rangle_{|\nu |} \simeq - \frac{1}{V} \chi_{t} + 
\frac{\nu^{2}}{V^{2}} + O(V^{-3}) \ .
\ee
For tests in 2-flavour QCD we refer to Ref.\ \cite{chitQCDNf2}.
(The original formula even includes a correction for a possible
deviation from a Gaussian distribution of the topological
charges, which we neglect.)
In order to justify the assumptions in the derivation of this 
formula, we have to assume a large expectation value
$\langle \nu^{2} \rangle = V \chi_{t}$, and a small ratio
$|\nu | / \langle \nu^{2} \rangle$.

As an example, we show in Fig.\ \ref{rhotcorre} the
corresponding correlation at $L=16$, $\nu =0$ and various
masses. Numerically the density was computed from the simplest
lattice version of $\rho_{t} = \epsilon_{12} F_{12}$ (this is
not problematic in the current setting, where we are always 
dealing with smooth configurations).
\begin{figure}[h!]
\begin{center}
\includegraphics[width=13pc,angle=270]{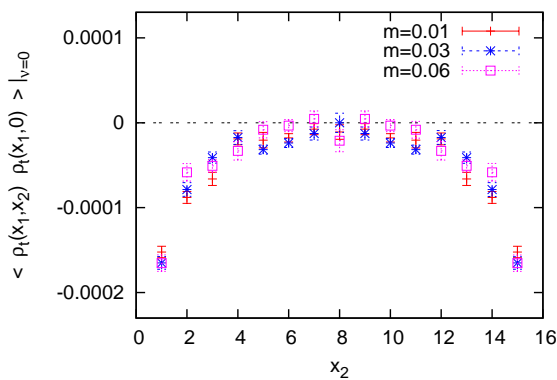}
\end{center}
\caption{\label{rhotcorre} The correlation of the topological
charge density for $L=16$, $\nu =0$ and $m=0.01,\ 0.03,\ 0.06$.
The statistical noise does not allow us to resolve 
a plateau value at large distances --- overcoming this problem 
would require a huge statistics.}
\end{figure}

At large distances one should find a plateau value,
which would then yield $\chi_{t}$. In particular for $m=0.06$,
where the maximal distance might be sufficient to see the asymptotic 
behaviour, we expect (based on the data and the conjectured formula
in Section 6) a plateau value of $\chi_{t}/V \approx
- 6 \cdot 10^{-6}$.
However, our statistical errors are of $O(10^{-5})$,
so in order to clearly resolve this plateau we would need
about $50 \, 000$ to $100\, 000$ configurations (cf.\ Table 
\ref{statab}). We conclude that the applicability of this 
method requires unfortunately a very large statistics.

\section{Approximate topological summation of observables}
\

We now proceed to the main approach in this study. It is a method
that does not require a known input observable either (as in Section 7),
but measurements in various topological sectors and volumes are 
needed. In fact this is the input which is usually accessible.
Then one tries to extract a (topologically summed) observable
$\langle O \rangle$ by employing the approximation formula
\be  \label{approsum}
\langle O \rangle_{|\nu |} \approx \langle O \rangle
+ \frac{c}{V \chi_{t}} \left( 1 - \frac{\nu^{2}}{V \chi_{t}}
\right) \ \qquad (c = {\rm const.}) \ .
\ee
This formula has been derived first for the pion mass in QCD
\cite{BCNW}, but it applies generally to observables in a
field theory with topology \cite{BHSV}. As in Section 7 
one assumes a Gaussian distribution of the topological charges, and 
a large value of $V \chi_{t}$, as well as a small ratio $|\nu | / 
\langle \nu^{2} \rangle$, are favourable for the validity of 
the approximations involved in the derivation.
This approximation formula could be truly powerful in QCD
and elsewhere, but it has never been tested before.

\subsection{Application to the chiral condensate}
\

Let us apply formula (\ref{approsum}) to the chiral condensate.
It is convenient to modify the notation,
\be  \label{AB}
\Sigma_{\nu} \approx \Sigma - \frac{A}{V} + \nu^{2} \frac{B}{V^{2}} \ ,
\qquad A = - \frac{c}{\chi_{t}} \ , \quad  B = - \frac{c}{\chi_{t}^{2}} \ .
\ee
The unknown quantities are $\Sigma$, $A$ and $B$, and we are
ultimately interested in $\Sigma$ and $\chi_{t} = A/B$.
They can be determined (in the framework of this approximation) 
by numerical results for some $\Sigma_{\nu}$:
\begin{itemize}
\item At fixed $m$ and $V$, we can determine $B$, for instance
from $\Sigma_{0}$ and $\Sigma_{1}$.
\item If we keep $m$ fixed but consider two volumes, $V_{1} \neq
V_{2}$, we can further determine $A$, {\it e.g.}\ based on $\Sigma_{0}$.
\end{itemize}
In total, it takes (at least) three $\Sigma_{\nu}$ values, involving 
two volumes, to obtain results for $\Sigma$ and $\chi_{t}$.

We follow this sequence of steps and start with the determination
of $B$. If we use as our input the measurements in the topological
sectors with $|\nu | = k, \ \ell$ (at fixed $m$ and $V$), we denote 
the result as $B_{k, \ell}$,
\be \label{Bkl}
\frac{1}{V} B_{k,\ell} = V \frac{\Sigma_{k}-\Sigma_{\ell}}{k^{2}-\ell^{2}}
= \frac{1}{m (k + \ell )} + V \frac{\varepsilon_{k}- \varepsilon_{\ell}}
{k^{2}-\ell^{2}} \ .
\ee
The semi-classical term, $1/(m(k+\ell ))$, tends to vary strongly 
for different choices of $k$ and $\ell$. Ideally the quantum effects
should render the results for $B_{k, \ell}$ similar again.
As an example, we show in Fig.\ \ref{Bk0} results for $B_{k,0}$
at $L=16$, $m=24$. In fact the non-perturbative results are much
more stable in $k$ than the semi-classical contributions alone.
Hence the first consistency test is passed well.\\
\begin{figure}[h!]
\begin{center}
\includegraphics[width=13pc,angle=270]{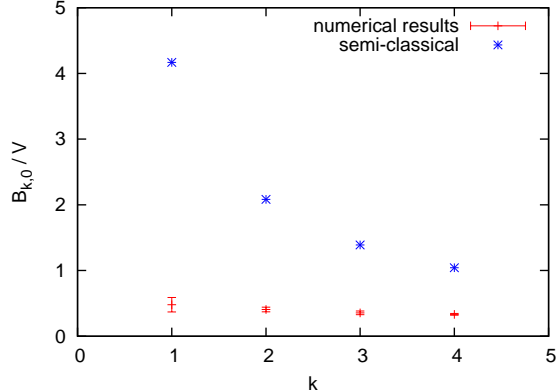}
\end{center}
\caption{\label{Bk0} Results for the auxiliary variables
$B_{k,0}$ (cf.\ eqs.\ (\ref{AB}), (\ref{Bkl})) at $L=16$,
$m=0.24$, for $k= 1, \dots ,4$. We see that the numerical results 
are quite consistent, in contrast to the semi-classical values.}
\end{figure}

We proceed to the determination of $A$, and therefore of $\Sigma$,
based on $\Sigma_{0}$ measurements in two volumes with sizes 
$(L_{1}, L_{2})$. Here we consider $m=0.01$ and we give two
examples:
\begin{itemize}
\item The $\Sigma_{0}$ values in $(L_{1}, L_{2})= (16,32)$ yield \
$\Sigma = 0.0199(7)$.
\item The $\Sigma_{0}$ values in $(L_{1}, L_{2})= (20,32)$ yield \
$\Sigma = 0.0207(12)$.
\end{itemize}
Thus the consistency looks fine again, but these results are far below
the prediction (\ref{Sigmatheo}), $\Sigma = 0.04888$ (in an infinite 
volume). In this case, our results are strongly affected
by finite size effects, which is not surprising: for the
given fermion mass, the correlation length in infinite volume
(given by eq.\ (\ref{Mpitheo})) would be 
$\xi = 1/M_{\pi, V=\infty} \simeq 14$. The relatively small boxes
enhance the Dirac eigenvalues $|\lambda_{i}|$, such that
$\Sigma$ decreases.\\

So the mass $m=0.06$ should be more promising, where theory
predicts $\xi \simeq 4.3$. Here we only have data in 
 $(L_{1}, L_{2})= (16,32)$, for $| \nu | = 0, \ 1$, so we cannot
repeat the above consistency tests. Nevertheless we can evaluate
$\chi_{t} = 0.00118(30)$ (which is just compatible with the
conjecture (\ref{Durrconj}), $\chi_{t} \simeq 0.00146$).
We further insert our most reliable result for $B$, namely
$B_{1,0}$ measured in $L=32$, and arrive at a result for $\Sigma$,
which is indeed close to the theoretical prediction (\ref{Mpitheo}),
\be
\Sigma_{\rm numerical} = 0.0883(69) \quad , \quad
\Sigma_{\rm theory} = 0.0888 \ .
\ee

\subsection{Application to the pion mass}
\

Let us also test the approximate summation formula (\ref{approsum})
by applying it to the pion mass. As we mentioned before, this 
was the original idea of Ref.\ \cite{BCNW} (though that work referred
to QCD). We re-write approximation (\ref{approsum}) in the notation 
analogous to (\ref{AB}),
\be
M_{\pi , |\nu |} \approx M_{\pi} - \frac{A}{V} + \frac{B}{V^{2}} \nu^{2} \ .
\ee
However, we now adopt a strategy which differs from the
previous consideration of $\Sigma$: at fixed $m$ we determine the 
three unknown parameters $A, \ B, \ M_{\pi}$ directly by a least-square
fit for some set of numerical $M_{\pi , |\nu |}$ values.\\

$\bullet$ For $m=0.01$ we have in total 11 measurements of
$M_{\pi , |\nu |}$ (see Table \ref{statab}), 
and we include the most promising ones.
We need at least two volumes, so we take the largest two
with $(L_{1},L_{2}) = (28,32)$. Moreover we only include the
topological sectors with $|\nu | \leq 1$, which are favourable
for the condition that $|\nu | / \langle \nu^{2} \rangle$
should be small. This leads to
\be
\left.
\begin{array}{ccc}
 & M_{\pi ,0} & M_{\pi ,1} \\
L=28: &         & 0.146(4) \\
L=32: & 0.05(1) & 0.160(8)
\end{array}
\right\} \ \ \overbrace{\longrightarrow}^{\mbox{fit}}
\ \  M_{\pi} =0.073(25) \ ,
\ee
which matches well the theoretical prediction, $M_{\pi} =0.071$ 
(albeit with a large error).\\

$\bullet$ We proceed to $m=0.06$, where we only have data for
$(L_{1},L_{2}) = (16,32)$. Hence we have less choice in this case,
but the finite size effects are less severe. Again we include
the results for $|\nu | \leq 1$, which corresponds to four
input measurements this time, and we arrive at
\be
\left.
\begin{array}{ccc}
 & M_{\pi ,0} & M_{\pi ,1} \\
L=16: & 0.041(1) & 0.271(4) \\
L=32: & 0.23(1)  & 0.232(7)
\end{array}
\right\} \ \ \overbrace{\longrightarrow}^{\mbox{fit}}
\ \  M_{\pi} =0.233(8) \ .
\ee
Also this result agrees well with the theoretical pion mass, 
$M_{\pi} =0.235$, and this time also the uncertainty is modest.

\section{Conclusions}
\

We have addressed a quite generic problem of lattice simulations in 
gauge theories with dynamical (quasi-)chiral fermions. The Monte Carlo
histories of such simulations tend to get trapped in {\em one} topological
sector for a very long (simulation) time, {\it i.e.}\ over many update
steps of the lattice gauge configuration. A conceptual issue that one 
has to address in this situation is {\em ergodicity}, a property which 
is compulsory for a correct algorithm. Here we studied a more practical
question: how can we evaluate the expectation value of some observable
$\langle O \rangle$, when only numerical measurements restricted
to a few topological sectors, $\langle O \rangle_{\nu}$, are available? \\

The dominant subject in contemporary
lattice simulations is QCD with dynamical quarks.
Here the problem of topological restriction is most striking when
one deals with chiral lattice quarks (of overlap \cite{alt} or 
Domain Wall \cite{Kaplan} type), which solve the Ginsparg-Wilson Relation 
(eq.\ (\ref{GWR}) or generalisations thereof). The use of Wilson type 
quarks is more widespread because they are much faster to
simulate, though plagued by additive mass renormalisation and
problems related to operator mixing.\footnote{Also that problem is 
avoided by the use of Ginsparg-Wilson fermions \cite{Has}.}
\footnote{For completeness we add that ``staggered fermions''
are widespread as well in lattice QCD. They are also quick to
simulate, and they do not suffer from additive mass renormalisation,
but the number of flavours is not flexible. Therefore it is now 
popular to take the fourth root of the fermion determinant
(cf.\ eq.\ (\ref{fermidet})), which formally corresponds to a
single flavour, but this is harmful for locality, which is
conceptually important. The question if this is a reason to
worry in practice is highly controversial. In any case,
neither Wilson nor staggered fermions do provide a sound
definition of the topological charge since there is no
well-defined fermion index, in contrast to Ginsparg-Wilson
fermions \cite{HLN}. Hence one has to refer to some rather
hand-waving definition in these cases.}
  
Here the aforementioned topological problem is less severe so far, 
but it is expected to show up as well when simulations will be
carried out on finer and finer lattices, say with lattice spacing 
$a < 0.05~{\rm fm}$. This renders the lattice QCD formulation
more and more continuum-like, which is in general welcome, but
it also makes it more difficult to change the topological sector.

This problem is not manifest in a very large volume, where 
$\langle O \rangle_{\nu}$ is the same for all indices $\nu$
(this property agrees with approximations (\ref{densecorr}),
(\ref{approsum})).
However, to suppress the topological dependence and other
finite size effects, the volume has to be
large compared to the correlation length, which is given by
the inverse pion mass, $\xi_{\rm QCD} \approx 1.4 ~{\rm fm} \ll 
La$. But when $a$ is very small, this requires a huge lattice size
$L$, which makes simulations again very tedious.

As a way out, the use of open boundary conditions in the Euclidean
time direction has recently been advocated, so that topological 
charge can gradually flow in or out of the volume during a simulation 
\cite{openbc}. In our study, however, we stay with periodic boundary 
conditions for the gauge fields, which guarantee that the topological 
charge is always integer, along with (discrete) translation invariance.
As a toy model we considered the Schwinger model with two
light, degenerate flavours, which were represented on the lattice
by dynamical overlap hypercube fermions. In a set of small
or moderate volumes, 
this only enabled measurements inside some specific topological 
sectors. In order to establish a link to the ``physical'' quantities,
we tested three methods to approximate the topological summation:\\

\begin{itemize}

\item The confrontation of a Gaussian summation with a known
observable allows us to fix the topological susceptibility $\chi_{t}$.
This method is robust, but it requires a known
input quantity. This is available in the 2-flavour Schwinger model
\cite{Smilga} (we used the chiral condensate), but not in general.\\

\item Next we tested a method to evaluate $\chi_{t}$ based on the
correlation function of the topological charge density \cite{AFHO}. 
More precisely, one searches for an asymptotic plateau of this correlation 
at large distances, which should amount to $-\chi_{t}/V$ (at $\nu =0$).
Unfortunately this value tends to be tiny for realistic settings, 
hence its resolution
requires a very large statistics.\\

\item Our main goal was the test of an approximate summation
formula given in Ref.\ \cite{BCNW}, which could provide a 
``physical'' result $\langle O \rangle$,
using only measurements of some topologically restricted observables
$\langle O \rangle_{\nu}$ as an input --- for various values of
$|\nu |$, in at least two volumes. This method is potentially
powerful, but it has never been tested before.

Our results suggest that it may work, if the assumptions used
in the derivation of this formula are reasonably well justified.
In particular, $V \chi_{t} = \langle \nu^{2} \rangle$ should be 
``large'', but it is difficult to predict explicitly what this
means. In our settings this quantity was always below 
$0.5$, but nevertheless we found decent (though not very precise)
results for the topologically summed chiral condensate and
pion mass. This observation is encouraging for applications
in QCD simulations with dynamical quarks.

\end{itemize}
\ \\

\noindent
{\bf Acknowledgements:} \ 
Stanislav Shcheredin and Jan Volkholz have contributed 
to this work at an early stage. We also thank Poul Damgaard,
Stephan D\"{u}rr, Hidenori Fukaya and Jim Hetrick for helpful comments.
This work was supported by the Croatian Ministry of Science, Education
and Sports (project 0160013) and by the Deutsche 
Forschungsgemeinschaft through Sonderforschungsbereich Transregio
55 (SFB/TR55) ``Hadron Physics from Lattice QCD''.

\end{document}